\documentclass[preprint,showpacs,pre,preprintnumbers,amsmath,amssymb]{revtex4}

\usepackage{graphicx}
\usepackage{dcolumn}
\usepackage{bm}

\begin{document} 
\title{Cage diffusion in liquid mercury}
\author{Yaspal S. Badyal}
\address{Oak Ridge National Laboratory, Oak Ridge, TN 37831}
\author{Ubaldo Bafile}
\address{Instituto di Fisica Applicata "Nello Carrara", Consiglio Nazionale delle Ricerche, Firenze, Italy}
\author{Kunimasa Miyazaki}
\address{ Department of Chemistry and Chemical Biology,
 Harvard University, MA 02138}
\author{Ignatz M. de Schepper}
\address{Interfaculty Reactor Institute, TU Delft, 2629 JB Delft, The Netherlands}
\author{Wouter Montfrooij$^*$}
\address{University of Missouri, Columbia, MO 65211}

\begin{abstract} 
{We present inelastic neutron scattering measurements on liquid mercury at room temperature for wave numbers $q$ in the range 0.3 $< q <$ 7.0 \AA$^{-1}$.
We find that the energy halfwidth of the incoherent part of the dynamic structure factor $S(q,E)$ is determinded by
a self-diffusion process. The halfwidth of the coherent part
of $S(q, E)$ shows the characteristic behavior expected for a cage diffusion process. We also show that the response function at small wave numbers exhibits a quasi-elastic mode with a time scale characteristic of cage diffusion, however, its intensity is larger by an order of magnitude than what would be expected for cage diffusion. We speculate on a scenario in which the intensity of the cage diffusion mode at small wave numbers is amplified through a valence fluctuation mechanism.} 
\end{abstract}
\maketitle
\section{Introduction}

The collective microscopic dynamics of dense hard-sphere fluids can be understood on the basis of cage diffusion.
In this process, the particles with hard sphere diameter
$\sigma_{hs}$, while all moving diffusively, find themselves locked up in
a cage formed by their nearest neighbors.
As a consequence, the coherent (collective) dynamic structure factor,
$S_{c}(q,E)$  of the system becomes, as a function of wave number
$q$ and energy $E$, very narrow in energy for wave numbers near
$q\sigma_{hs}\simeq 2\pi$, i.e., near the peak in the static structure factor $S(q)$. This narrowing is referred to as de Gennes narrowing of
$S_{c}(q,E)$. It is believed that cage diffusion plays an important part in the dynamics of real fluids, such as noble gas fluids\cite{peter,cage},
concentrated colloidal suspensions\cite{pusey,schepper} and liquid metals\cite{liviaprl,livia}.\\
Over the past decennia, the microscopic dynamics of liquid metals have been the subject of intensive study by neutron scattering and X-ray techniques\cite{ru,na,na2,na3,cs,li2,li3,k,k2,hg,ge}, and molecular dynamics simulations\cite{livia,li,na4}. Liquid metals pose a particularly interesting problem because of the long range of the interatomic potential. The (extended) hydrodynamics modes have been investigated in these systems, and the influence of the mode coupling mechanism on these modes\cite{na2,na3}, leading to long-time tails in the correlation functions. In here, we present results on the fast (in time) decay mechanism in liquid mercury pertinent to cage diffusion, as first investigated by Bove {\it et al.}\cite{liviaprl}.\\
Recent neutron scattering experiments by Bove {\it et al.}\cite{liviaprl} on liquid mercury at room temperature showed that the incoherent part of the scattering function (i.e., the dynamics of individual Hg atoms) was characterized by two time scales. The authors attributed the presence of more than one time scale to the process of cage diffusion. These two time scales were also present in Molecular Dynamics (MD) simulations carried out by the same group\cite{livia}. However, their neutron scattering setup did not allow for a detailed investigation of the incoherent dynamic response, as it was setup primarily to investigate $q$-dependence of the sound modes outside the hydrodynamic regime.\\ 
We report inelastic neutron scattering experiments on Hg at room temperature (number density 0.0408 \AA$^{-3}$) at a higher energy resolution, allowing us to investigate whether cage diffusion is indeed present in Mercury, and determine the origin of the two time scales found previously\cite{livia}. Our results show that cage diffusion is the dominant dynamic process in Hg. We find that the main contribution to the incoherent cross section is determined by a self-diffusion process. The second time scale is manifest in the neutron scattering response as a broad quasi-elastic line with an energy halfwidth of 2 meV. The mode associated with this second time scale disappears from the spectra for $q >$ 1 \AA$^{-1}$, however, its intensity at the smallest $q$-values is an order of magnitude larger than what could be expected from cage diffusion. We show that this mode cannot be viewed as an extension of the hydrodynamic Rayleigh mode to finite $q$-values. Instead, we propose that the intensity of the cage diffusion mode in the neutron scattering spectra at small wave numbers has been amplified through a valence fluctuation mechanism. Other than the presence of this broad quasi-elastic contribution, we find that the response of liquid mercury can be described in terms similar to a range of liquids investigated previously, albeit that Hg has a very large anomalous sound dispersion reflecting the presence of the electron gas\cite{liviaprl}.\\

\section{Experiments and Data Analysis}

The inelastic neutron scattering experiments were performed on natural mercury, using the time-of-flight spectrometer MARI at the Rutherford Appleton Laboratories. Mercury scatters neutrons both coherently and incoherently, yielding information on both the collective and individual dynamics of the Hg atoms, respectively. Unfortunately, natural-enrichment mercury also contains neutron absorbing isotopes, thus rendering the data correction procedure more cumbersome than usual.\\
In order to minimize neutron absorption, we used a thin disk container of 0.08 cm nominal internal thickness, and 11.2 cm radius. Niobium disks of 0.12 cm thickness were used for the walls of the container. Two sets of measurements were taken on MARI with different incident neutron energies, 20 meV and 40 meV, with the sample cell oriented perpendicular to the beam, and at an angle of 45$^o$, respectively. For each incident energy, we collected sample spectra, empty container spectra on a separate container of identical dimensions, vanadium spectra on an vanadium disk of 0.084 cm thickness in the same orientation as the sample, and we measured the neutron transmission ratio for each of these measurements. The latter allowed us to determine the actual thickness of the mercury in the beam by comparing it to the empty container transmission. We found the thickness to be 0.0816 $\pm$ 0.0004 cm, and the orientation for the 40 meV measurements to be 44$^o$.8 $\pm$ 0$^o$.6. In our correction procedure, we use these measured values for calculating the attenuation of the sample. All measurements were taken under ambient conditions.\\
The neutron scattering experiments were carried out in 1995 and were originally setup to measure the sound dispersion in Hg. Unfortunately, we underestimated the propagation velocity of the extended sound modes, and we failed to observe a clear signal associated with the anomalous sound dispersion\cite{liviaprl} as it mostly lay outside of our kinematic scattering window. However, we could determine the halfwidths of the coherent and incoherent part of the scattering function for wave numbers $q$ with 0.3 $< q < $ 7 \AA$^{-1}$, the results for which we report in this paper.\\
For the data reduction procedure, we first subtracted the time-independent background, obtained from an empty spectrometer run, and corrected the data for the detector efficiencies, as obtained from a vanadium reference sample. Then, we multiploied the empty container spectra by multiplying them with the energy and angle dependent sample attenuation factor for Hg\cite{sears}. These sample attenuation factors are typically of the order of 5-20 for our highly-absorbing Hg sample. Next, the attenuated container spectra were subtracted from the sample plus container spectra, and the results were divided by the calculated\cite{sears} attenuation factor of Hg. Finally, the results were corrected for multiple scattering, a 5-10\% effect, by carrying out a numerical integration over the sample volume using the measured\cite{ubaldo} static structure factor as input and assuming a Lorentzian lineshape for the dynamic structure factor with halfwidth given by the mercury self-diffusion coefficient (0.1436 \AA$^2$/ps at 283 K\cite{lobo}).\\ 
As a check on our data reduction procedure, we show in Fig. [\ref{attenuated}] the measured 20 meV data, integrated over energy($|E|$ $<$ 4 meV) as a function of scattering angle after subtraction of container scattering, but before the sample attenuation correction. In this figure, the dip at 90 degrees is due to the large absorption cross section of Hg and the flat plate design of the sample cell, so that the sample effectively shields itself at these scattering angles. The width of the dip represents Cd shielding on the side of the container. We compare the energy integrated results to the calculated total (single plus double) scattering obtained using the published static structure factor\cite{ubaldo} and diffusion constants\cite{lobo,handbook}, multiplied with the calculated attenuation factor. It is clear from this figure that the agreement is good at least to within 10\%, allowing for an accurate determination of the halfwidths of the spectra. For our analysis, we only use the spectra with $\phi$ $<$ 80$^0$ for the 20 meV dataset, and $\phi$ $<$ 120$^0$ for the 40 meV dataset. We show the fully corrected data for both incident neutron energies as a contour plot in Fig. [\ref{corrected}]. Close inspection of this figure shows the presence of a spurion at $E$ = 5 meV energy transfer for the 20 meV data set, and at $E$ = 10 meV for the 40 meV data set. This spurion corresponds to neutrons that are scattered elastically, but have traveled an additional 60 cm. We believe that this corresponds to neutrons that were scattered by the sample, and subsequently scattered by the flange holding the sample in position. In the 1995 MARI setup, this flange was insufficiently shielded. For our data analysis, we ignored the region around the spurion. In addition, the corrected data show small remnants due to the Bragg peaks of the Nb container (such as at $q$ = 4.7\AA$^{-1}$ in Fig. [\ref{corrected}]). Affected $q$-values were also ignored in our data analysis.\\
In order to determine the coherent and incoherent halfwidths from the neutron scattering spectra, we fit the data to a number of Lorentzian lines in a procedure similar to the one reported by Bove {\it et al.}\cite{liviaprl}. We find that a two Lorentzian line description gives good agreement ($\chi^2 \sim$ 1) with our measurements $q <$ 1.9 \AA$^{-1}$. We therefore fit these spectra to
\begin{equation}
S_{tot}(q,E) = \frac{sE}{1-e^{-E/k_BT}} [\frac{\sigma_i}{\sigma_c}\frac{\Gamma_i(q)}{E^2+\Gamma_i^2(q)}+\frac{\chi(q)\Gamma_c(q)}{E^2+\Gamma_c^2(q)}]. 
\label{fitfunction}
\end{equation}
In here, $k_B$ is Boltzmann's constant, $\sigma_i$ and $\sigma_c$ are the incoherent and coherent cross sections for Hg ($\sigma_i$/$\sigma_c$= 0.324), $\Gamma_i(q)$ is the incoherent halfwidth at half maximum (HWHM),  
$\Gamma_c(q)$ is the coherent halfwidth, and $\chi(q)$ is the static susceptibility. At room temperature, $\chi(q)$ is almost identical to the static structure factor $S(q)$. The prefactor $s$ takes the minor difference between $\chi(q)$ and $S(q)$ into account, as well as any remaining scale factor due to the attenuation correction. The fit function was convoluted with the experimental resolution function obtained from the vanadium measurements, which is well described by an Ikeda-Carpenter form. The full width at half maximum in energy for neutrons scattered elastically for the 20 meV (40 meV) measurement is 0.4 meV (0.8 meV), independent of scattering angle (cf. Fig. [\ref{lowq}]).\\
For $q >$ 1.9 \AA$^{-1}$, we find that we obtain a better fit if we describe the coherent part of the scattering by multiple Lorentzian lines. We find that two Lorentzian lines (damped harmonic oscillator\cite{wouter}) give a good description at most $q$-values, but that around the first peak of the static structure factor the quality of the fit can be improved further by fitting to three Lorentzian lines (visco-elastic model\cite{lovesey}). The (coherent part of the) damped harmonic oscillator model has one free parameter, the visco-elastic model has two, and both models satisfy the f-sum rule.\\
Note that we only use the fit function to reliably determine the halfwidths of the coherent and incoherent parts of the spectra, corrected for the spectrometer resolution function. We do not address the individual fit parameters as they come out of the visco-elastic or harmonic oscillator models; we find that the fit parameters in the visco-elastic model are highly correlated due to the presence of both coherent and incoherent scattering. However, Eq. [\ref{fitfunction}] and modifications thereof give an overall good description of the data, and the derived $\Gamma$'s should be an accurate representation of the halfwidths.\\ 
Examples of the fits are shown in Figs. [\ref{lowq}] and [\ref{fits}]. At low $q$, the spectra are clearly a combination of a narrow contribution, and a broad contribution. We verified that this broad contribution was not an artifact of our correction procedure by noting that it was completely absent in the vanadium reference spectra (see Fig. [\ref{lowq}]), nor could it be attributed to container scattering. At low $q$, the incoherent line width is so small that the incoherent scattering is identical to the (asymmetric) MARI resolution function. It is only possible to determine $\Gamma_i(q)$ for $q > 1.5 $\AA$^{-1}$ where the incoherent halfwidth becomes comparable to the width of the energy resolution function. We fitted $\Gamma_i(q)$ for $1.5 < q < $ 2.1 \AA$^{-1}$, and for $q \approx$ 3.3 \AA$^{-1}$. In these $q$ regions the coherent and incoherent line widths are significantly different from eachother and allow for a reliable determiniation of the incoherent line width. We find that the incoherent line width is determined by the constant of self-diffusion $D_{s}$, with $D_{s}$ = 0.143 $\pm$ 0.007 \AA$^2$/ps. This value is slightly smaller than the expected value at room temperature (0.1436 at T= 283 K and 0.159 at T= 298 K\cite{lobo,handbook}). We did not have a temperature sensor on the sample while it was positioned inside the evacuated MARI spectrometer, nor did we log the temperature of the ISIS experimental hall during our October experiment. In all, our value seems to be reasonable, albeit on the low side of what is expected\cite{lobo,handbook}.\\
In order to determine the coherent line widths, we fixed the halfwidth $\Gamma_i(q)$ of the incoherent part at $D_{s}q^2$ with  $D_{s}$ = 0.143   \AA$^2$/ps, corrected for the expected decrease with $q$ as calculated from the Enskog theory\cite{enskog} (see Fig. [\ref{hw}b]). This decrease can be calculated without ambiguity, and represents the slow approach to ideal gas behavior at very large $q$. Having fixed the incoherent line widths, we then fitted the coherent part of the scattering to one, two or three Lorentzians. When we fitted to more than one Lorentzian, we fixed $\chi(q)$ to the value measured for $S(q)$ by neutron diffraction experiments\cite{ubaldo}. From the fit results, we then determined the coherent line width, the results for which we show in Fig. [\ref{hw}], together with the incoherent line width and the static structure factor\cite{ubaldo} for comparison.\\ 

\section{Discussion}

Inspecting Fig. [\ref{hw}a], one observes that for $q > $ 2 \AA$^{-1}$ the coherent halfwidth oscillates around the incoherent halfwidth, given by the coefficient for self-diffusion. We emphasize this oscillatory behavior at the higher momentum transfers by dividing the halfwidth by $q^2$ (see Fig. [\ref{hw}b]). From this figure we conclude that the oscillations of the coherent halfwidth are in phase with the oscillations of the static structure factor (see Fig. [\ref{hw}c]). In other words, whenever the static structure factor reaches a maximum (minimum), the coherent halfwidth reaches a minimum (maximum). This is what one expects\cite{cage} from cage diffusion: it should take longer to diffuse out of the cage formed by its nearest neighbors for wave numbers corresponding to a peak in the structure factor.\\
In Fig. [\ref{hw}] we also plot the predictions of the Enskog theory\cite{enskog} for hard spheres. Comparing the properties of a liquid to those of an equivalent hard-sphere fluid greatly facilitates the interpretation of experimental results, and it has been applied successfully to a host of liquids and gases\cite{host}. The one problem in this comparison lies in determining an equivalent hard-sphere diameter $\sigma_{hs}$, something which cannot be done unambiguously. For instance, comparing the height of the first maximum of $S(q)$ to that of the equivalent hard-sphere fluid yields $\sigma_{hs}$= 2.73 \AA, but comparing the position of this first maximum to that of the hard-sphere fluid gives $\sigma_{hs}$= 2.89 \AA, while comparing the density of the liquid at the melting point to the density at which a hard-sphere fluid melts\cite{vanloef} results in $\sigma_{hs}$= 2.81 \AA. Somewhat arbitrarily, we have opted for the latter method. This choice has the advantage that the calculated incoherent halfwidth from the Enskog theory\cite{enskog} coincides with our experimentally determined self-diffusion coefficient. The results for the coherent and incoherent halfwidths for the hard-sphere fluid are shown in Fig. [\ref{hw}a] and [\ref{hw}b]. The comparison with the static structure factor is shown in Fig. [\ref{hw}c].\\
The agreement in the phase of the oscillations between mercury and the hard-sphere fluid is rather good (dashed-dotted curves in Fig. [\ref{hw}]), but the amplitude of the oscillations is somewhat off, in particular at intermediate $q$-values. The situation improves when we use the measured $S(q)$ instead of the calculated hard-sphere $S(q)$ (solid curves in Fig. [\ref{hw}]). In all, from the qualitative agreement between the liquid mercury data and a hard-sphere fluid that is known to exhibit cage diffusion\cite{cage}, combined with the observation that the oscillations of the coherent halfwidth are in phase with those of the static structure factor, we conclude that cage diffusion is the dominant mechanism determining the dynamic response of liquid mercury at room temperature.\\
Next, we focus on the low-$q$ region ($q < 1 $\AA$^{-1}$) where Bove {\it et al.} have performed their neutron scattering experiments and MD computer simulations\cite{liviaprl,livia}. Bove {\it et al.} observed in their neutron scattering experiments the existence of a mode of characteristic half width at halh maximum (HWHM) of 2 meV, and an additional contribution to the single-particle dynamics (other than simple diffusion) in their MD simulations with a characteristic time scale of 0.3 ps. Since a time scale of $\sim$0.3 ps would show up as an energy width of
$\sim$ 2 meV in neutron scattering experiments, the authors suggested that the neutron scattering mode and the MD mode were the same, and identified this mode as being related to cage diffusion. We agree with the identification of the MD mode as being related to cage diffusion, but we argue that the 2 meV mode in the neutron scattering experiments cannot be fully explained by a straightforward cage diffusion mechanism.\\
In Fig. [\ref{lowq}] we show our experimental data for $q$= 0.45 \AA$^{-1}$. Data at $q$-values close to this value show a similar pattern: a resolution limited central peak, and a broader contribution centered around zero energy transfer with a characteristic HWHM of 2 meV for all $q$ with $q <$ 1 \AA$^{-1}$ (see Fig. [\ref{hw}a]). In addition, the absolute values of the intensities for the broad contribution agree well with those measured by Bove {\it et al.}\cite{liviaprl}.  In our fit, we had tentatively attributed this broad mode at small $q$-values to the coherent part of the scattering (see Eq. [\ref{fitfunction}]), however, we argue in the following that this is unlikely. There are two obvious candidates for the broad mode: the aforementioned cage diffusion mode seen in MD simulations, and the extension of the hydrodynamic Rayleigh mode (heat diffusion) to larger momentum transfers. We discuss both candidates.\\
First, it is expected that the neutron scattering spectra will show the equivalent of the hydrodynamic Rayleigh mode, a coherent process associated with heat diffusion. This mode is commonly referred to as the extended heat mode, hence our initial assignment of this broad mode to the coherent part of the scattering. In hydrodynamics, the expected intensity associated with the Rayleigh mode is $S(0)$($\gamma$-1)/$\gamma$, with the static structure factor $S(0)$ (= 0.005) at $q$ = 0 given by the compressibility, while $\gamma$ is the ratio of the specific heats (= 1.14 at T=283 K). Thus, the expected intensity of this heat mode (0.0006) compared to the expected intensity of the incoherent scattering ($\sigma_i$/$\sigma_c$= 0.324) is completely negligible in the hydrodynamic region. With increasing $q$, the heat-mode intensity will increase predominantly due to the increase in $S(q)$ and to a lesser extent due to a change in $\gamma(q)$. The halfwidth associated with the extended heat mode is shown as the dashed-dotted line in Fig. [\ref{hw}a] calculated using the Enskog theory\cite{enskog}. It is clear from this figure that for $q <$ 1 \AA$^{-1}$, that the measured halfwidth reaches a constant value, instead of decreasing $\sim q^2$ as expected for the extended heat mode. Moreover, the intensity of the 2 meV mode actually increases with decreasing $q$, an observation which also directly follows from Fig. [\ref{attenuated}]. Thus, the constant halfwidth of the 2 meV mode, its intensity ($\sim$ 0.07 at $q$ = 0.45\AA$^{-1}$) being far greater than what could be expected of the extended heat mode, combined with its increase in intensity with decreasing $q$ instead of the expected decrease, all lead us to conclude that the 2 meV mode is not an extended heat mode for $q <$ 1\AA$^{-1}$.\\ 
Second, if a particle is locked up in a cage, it will rattle around on short time scales before it escapes. Typical time scales correspond to how long it will take a particle to cover the distance to its neighbor (this distance is typically of the order of $\sigma_{hs}$/10) with an average thermal velocity $\sqrt{k_BT/m}$, $m$ being the mass of a Hg atom. For mercury, one can expect characteristic times very similar to those found in the MD simulations\cite{livia}. However, the influence of this process to the decay of the self-correlation function is very small at small $q$-values. The reason for this is that a particle just rattles around in its cage, never moving very far from its original position, so when it is probed with a long wavelength (small $q$) density fluctuation (such as a neutron scattering with small momentum transfer), only a very small effect is to be expected. The effect will be of the order 1-cos($q\Delta r$), with $\Delta r$ the distance between the surfaces of neighboring hard-spheres. For $q$= 0.3 \AA$^{-1}$ and $\Delta r$= 0.3 \AA, the expected effect compared to the standard self-diffusion mechanism will be less than 0.5\%. In the MD simulations\cite{livia}, the observed effect was 0.4$\%$. Based on these numbers, it seems entirely justifiable to ascribe the additional mode observed in MD simulations to cage diffusion. Conversely, the expected intensity in neutron scattering experiments should then also be 0.4\% of the intensity associated with the standard self-diffusion mode. However, when we compare the intensity of the broad mode to the resolution limited self-diffusion mode, we find a ratio in excess of 20 \% for the lowest $q$-values. This observation is at odds with the expected intensity ratio of $<$ 0.5\%, and we therefore rule out a straighforward cage diffusion mechanism as the dominant constituent of the broad mode at the lower $q$-values. We do not dispute the existence of a mode related to cage diffusion in the neutron scattering spectra, we just calculate that its predicted intensity is too small to account for the observed intensity.\\
We can only speculate as to the origin of the 2 meV mode in the neutron scattering spectra. First, this mode (see Fig. [\ref{hw}]) is not an artifact of our data correction procedure as our time-of-flight results are quantitatively similar to the neutron scattering experiments\cite{liviaprl} performed on a triple-axis spectrometer. Second, this mode corresponds to excess intensity compared to what is expected for coherent plus incoherent scattering by mercury (see Fig. [\ref{corrected}]). 
One possible explanation could be paramagnetic scattering by Hg ions. Whereas liquid mercury is diamagnetic, it is possible that valence fluctuations occur on very short time scales, as these time scales are not probed in a magnetic susceptibility experiment. If only a small fraction of the Hg-ions are not Hg$^{++}$, then a significant paramagnetic signal can be expected: 1.8 barns per magnetic ($S$= 1/2) atom compared to 6.5 barns for the incoherent scattering cross section of an Hg ion. Of course, the magnetic form factor for a paramagnetic liquid would drop off precipitously with increasing $q$, and the magnetic scattering intensity would rapidly diminish with increasing $q$.\\
Supposing that the agreement between the neutron scattering line width of 2 meV ($\sim$3 ps$^{-1}$) and the MD time scale of $\sim$1/3 ps\cite{livia} is more than a coincidence, we propose the following mechanism. When Hg ions aproach each other and move out again, as is the case with cage diffusion, then the overlap between the orbitals will change significantly. On close approach, the Coulomb repulsion could force one or more d-electrons into the sea of conduction electrons, whereas if a Hg ion happens to be relatively far removed from surrounding Hg ions, a non-bound electron could occupy the 4s orbital. Either of these processes would lead to a magnetic neutron scattering cross section, and these valence fluctuations would naturally occur on the same time scales as those of the cage diffusion process. This mechanism would therefore lead to an enhanced cross section for the cage diffusion process. Extrapolating the observed intensity of the 2 meV mode to $q$= 0, we find that the cross section for the 2 meV mode can be as high as $\sim$1.5 barns per Hg atom. This would correspond to the Hg ions spending significant part of their time with a valency different from 2+. This proposed scenario can easily be verified by future neutron scattering experiments: an increased temperature would lead to an even shorter time scale for the valence fluctuations, and therefore to an increased linewidth in energy.\\
Finally, we point out that somewhat similar effects have been observed in other liquid metals. For example, in liquid lithium it was observed\cite{li} that the measured static structure factor was larger than the calculated structure factor for $q <\sim$ 1 \AA$^{-1}$, in liquid cesium a larger than the literature value of $\sigma_i$ had to be used\cite{cs} in order to describe the neutron scattering data, and in liquid potassium it was found\cite{k2} that the quasi-elastic coherent line width at $q$ = 0.4 \AA$^{-1}$ had decreased by only $\sim$ 30 \% from its maximum value at $q \sim$ 1.4 \AA$^{-1}$. We are currently investigating whether these observations can be explained by a scenario similar to the one outlined in the preceding paragraphs.\\ 
In conclusion, we have shown that cage diffusion plays a dominant role in the collective dynamics of liquid mercury. This is reflected in the wave number dependence of the characteristic width of the coherent dynamic structure factor. The expected intensity in the incoherent dynamic structure factor from the signal attributable to cage diffusion is too small to account for the observed intensity at small wave numbers, however, it is possible that valence fluctuations on the time scale of cage diffusion greatly enhance the neutron scattering cross section for this process at the smaller wave numbers.\\
\begin{acknowledgments}
We are deeply indebted to the late Dr. Peter Verkerk, without whom this project never would have materialized. Also, we would like to thank Dr. L.E. Bove for very helpful discussions. The experiments at ISIS were supported by the Netherlands organization for Scientific Research (NWO), and part of this work is supported by grants from the
Japan Society for the Promotion of Science (JSPS).
\end{acknowledgments}

$^*$ Corresponding author.

\newpage
\begin{figure}
\caption[]
{The neutron scattering function $S(\phi)$ (circles) as a function of scattering angle $\phi$, integrated over energy transfers -4 $<E< $ 4 meV. The q-value corresponding to the elastic channel is shown for comparison at the top of the graph. The incident neutron energy was 20 meV, and the sample container was positioned perpendicularly to the neutron beam. The data in this figure have been corrected for container scattering, but not for attenuation. The solid line is the sum of the calculated single and double scattering (see text), and the dashed-dotted line is the contribution due to double scattering in the sample. Note the discrepancy at the smallest scattering angles between the calculated and measured intensity, indicative of an increased neutron scattering cross section. The sharp feature at 50 degrees is a remnant of container Bragg scattering.}
\label{attenuated}
\end{figure}
\begin{figure}
\caption[]
{The fully corrected dynamic structure factor $S(q,E)$ for Hg at room temperature. The contour plot (with the contour values given by the grey scale at the top of the figure) is a combination of the data sets with an incident neutron energy of 40 meV, and 20 meV. The latter corresponds to the higher energy resolution data for $q <$ 4 \AA$^{-1}$. Note the presence of spurions (see text) at $E$= 5 meV and 10 meV (shown by horizontal solid lines), and remnant Bragg scattering due to the Nb container near, for example, $q$= 4.7 \AA$^{-1}$.}
\label{corrected}
\end{figure}
\begin{figure}
\caption[]
{The solid circles are a cut through the fully corrected data at $q$= 0.45 \AA$^{-1}$, the solid line is the fit according to Eq. [\ref{fitfunction}]. The bottom figure is an enhancement of the top figure. The sharp component is given by the MARI resolution function, the broad \'2 meV\' component (with HWHM of 2 meV) is plotted separately as the dashed curve. For the sake of comparison, we have included the response function of vanadium (dotted line with open circles) measured under identical conditions and sample geometry, normalized to have the same intensity as the Hg data. Note that there is no indication of a broad contribution to the vanadium scattering, the scattering has fallen to zero for $E> $ 1 meV. The peaks near 5 meV are spurious.}
\label{lowq}
\end{figure}
\begin{figure}
\caption[]
{Examples of the results of the fitting procedure (solid lines) to the measured $S(q,E)$ (solid circles). The q-values are given in the figure. The peaks at 5 meV and 10 meV are spurious.}
\label{fits}
\end{figure}
\begin{figure}
\caption[]
{(a): The HWHM for the coherent part of $S(q,E)$ as a function of wave number $q$. The solid circles refer to a fit of the coherent part with two or more Lorentzian lines, the open circles denote the results of the fit with a single Lorentzian line. For 2.5 $< q <$ 3.5 \AA$^{-1}$ the results for the 20 meV and the 40 meV datasets overlap, both are shown. The three curves are calculated using the Enskog theory with $\sigma_{hs}$= 2.81 \AA. The dotted line is the incoherent HWHM, the solid curve is the prediction for the coherent HWHM using the calculated $S(q)$ for the equivalent hard-sphere fluid shown in part (c) (denoted hs in the legend), while the jittery dashed-dotted curve (denoted hs+$S_{exp}(q)$) is the prediction for the coherent HWHM using $S(q)$ from experiment\cite{ubaldo} (b): the same as in (a), except everything has been divided by $q^2$.\\
(c): The static structure factor $S(q)$ measured by neutron diffraction experiments\cite{ubaldo} (circles), and compared to $S(q)$ calculated using the Percus-Yevick theory with Henderson-Grundke correction factor \cite{py} for $\sigma_{hs}$= 2.81 \AA. (solid line)}
\label{hw}
\end{figure}

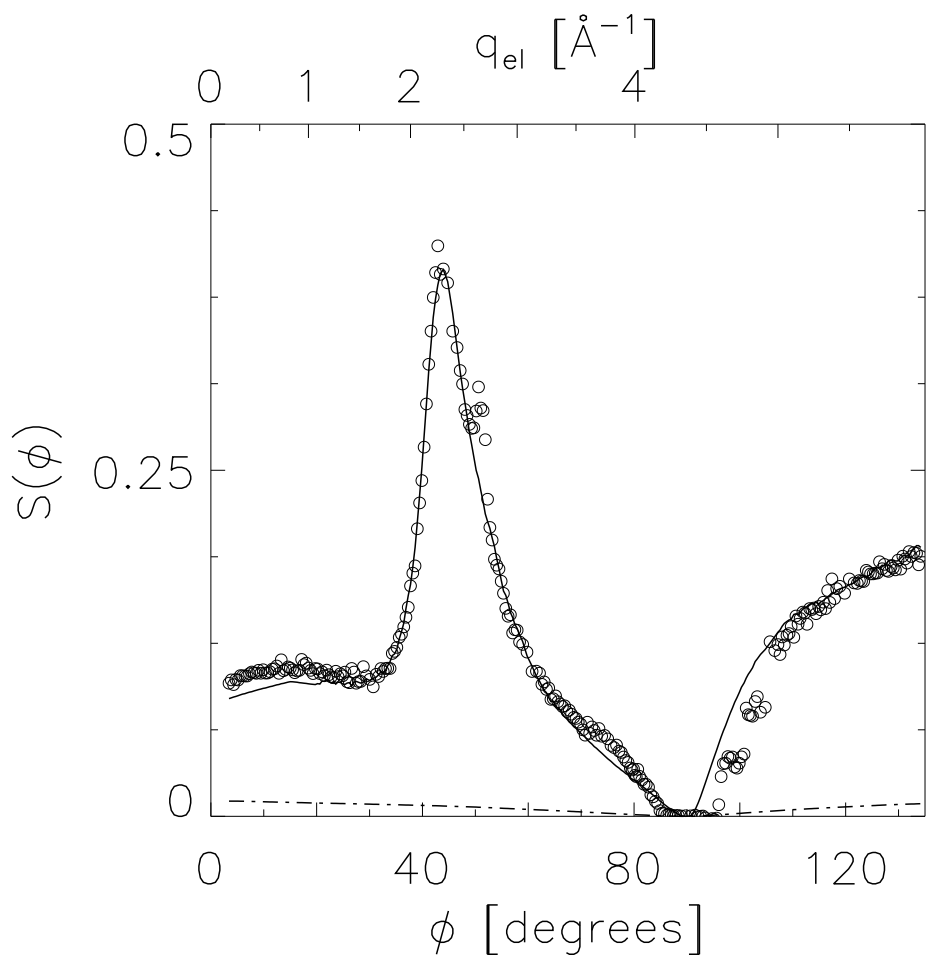
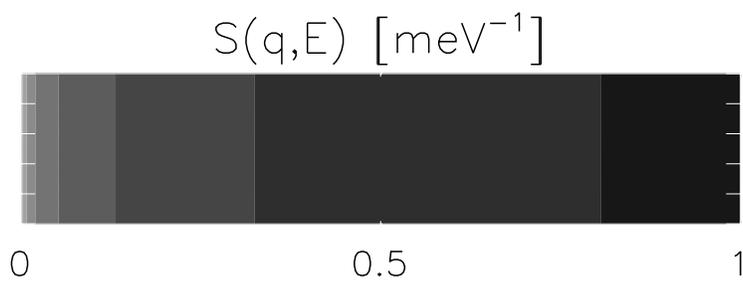
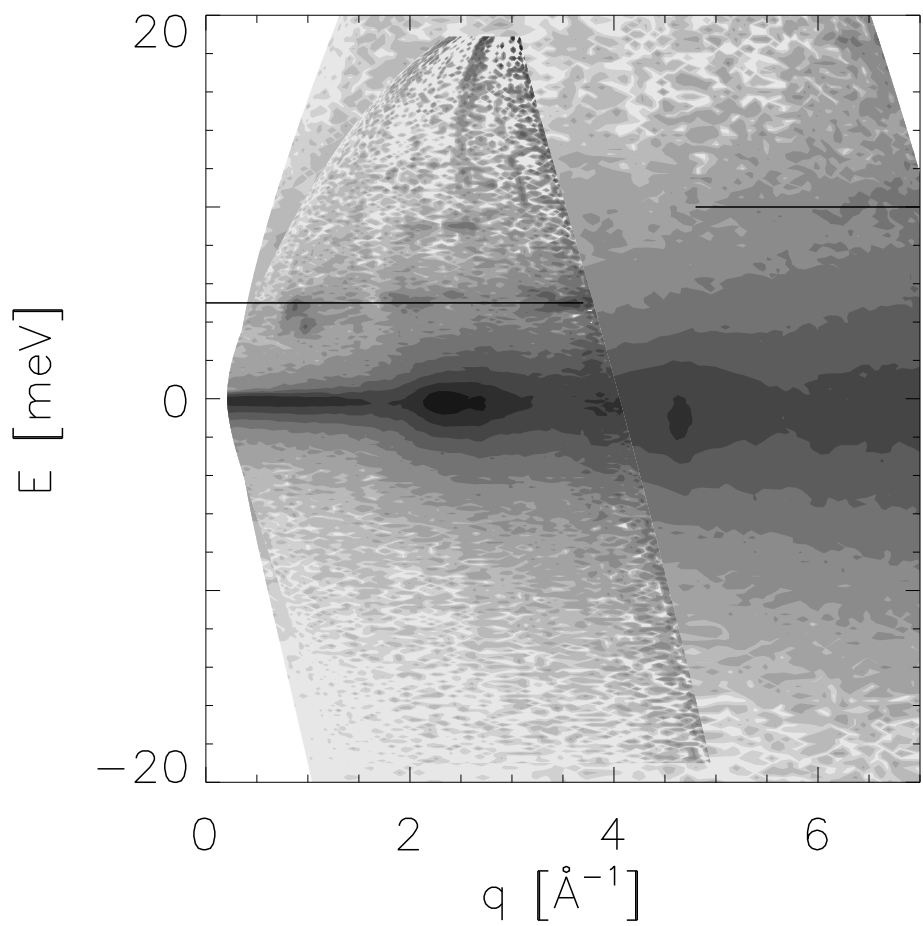
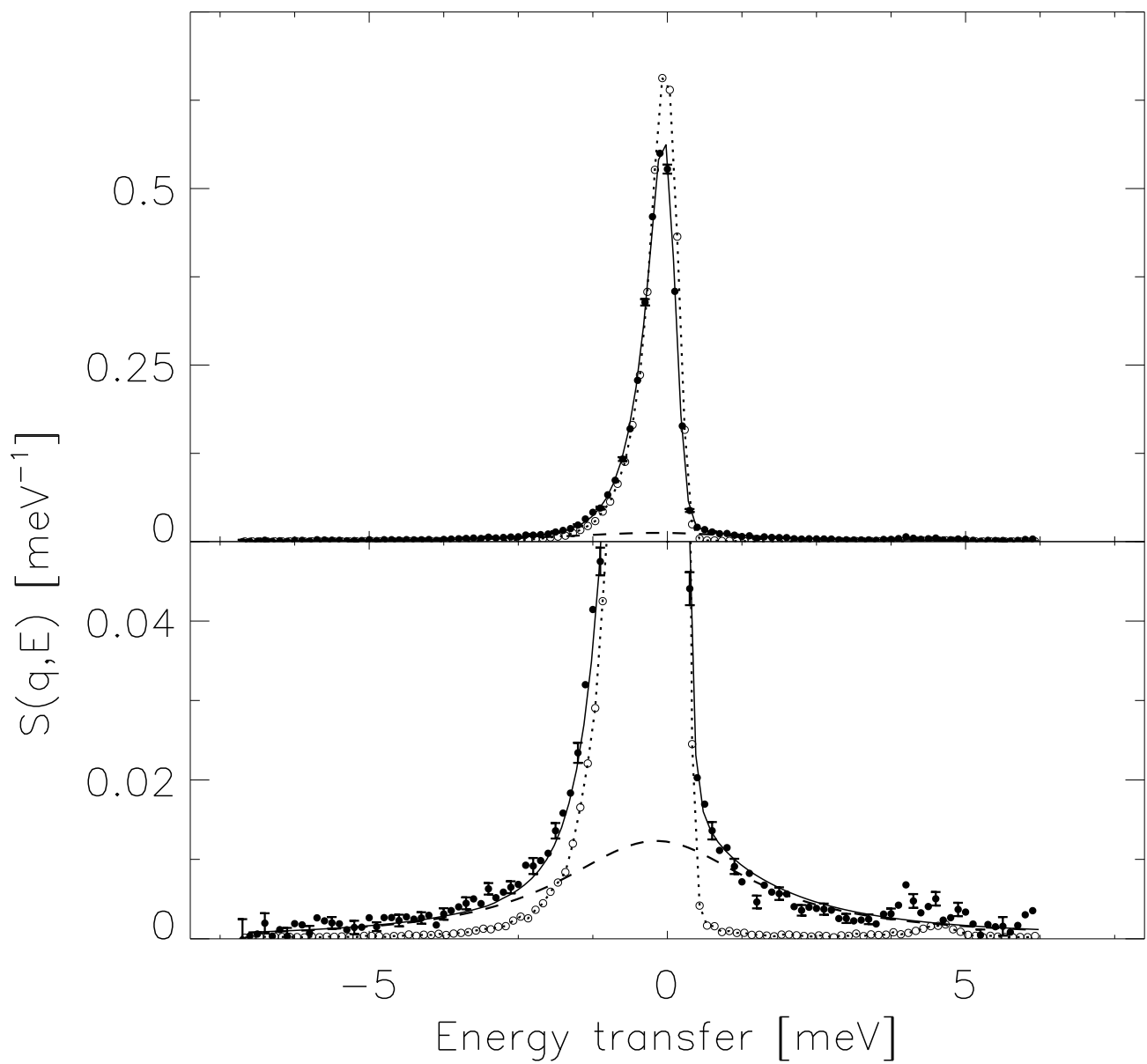
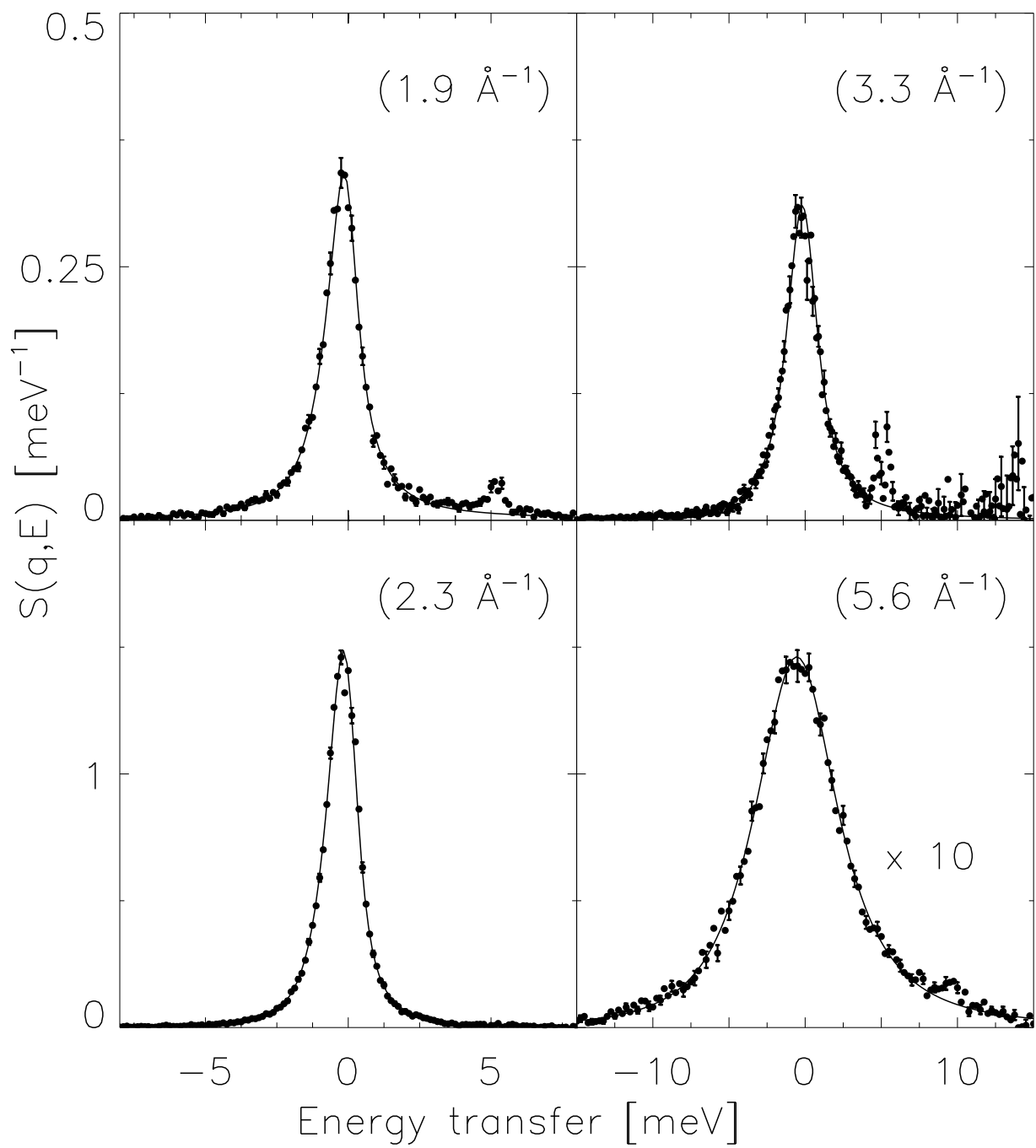
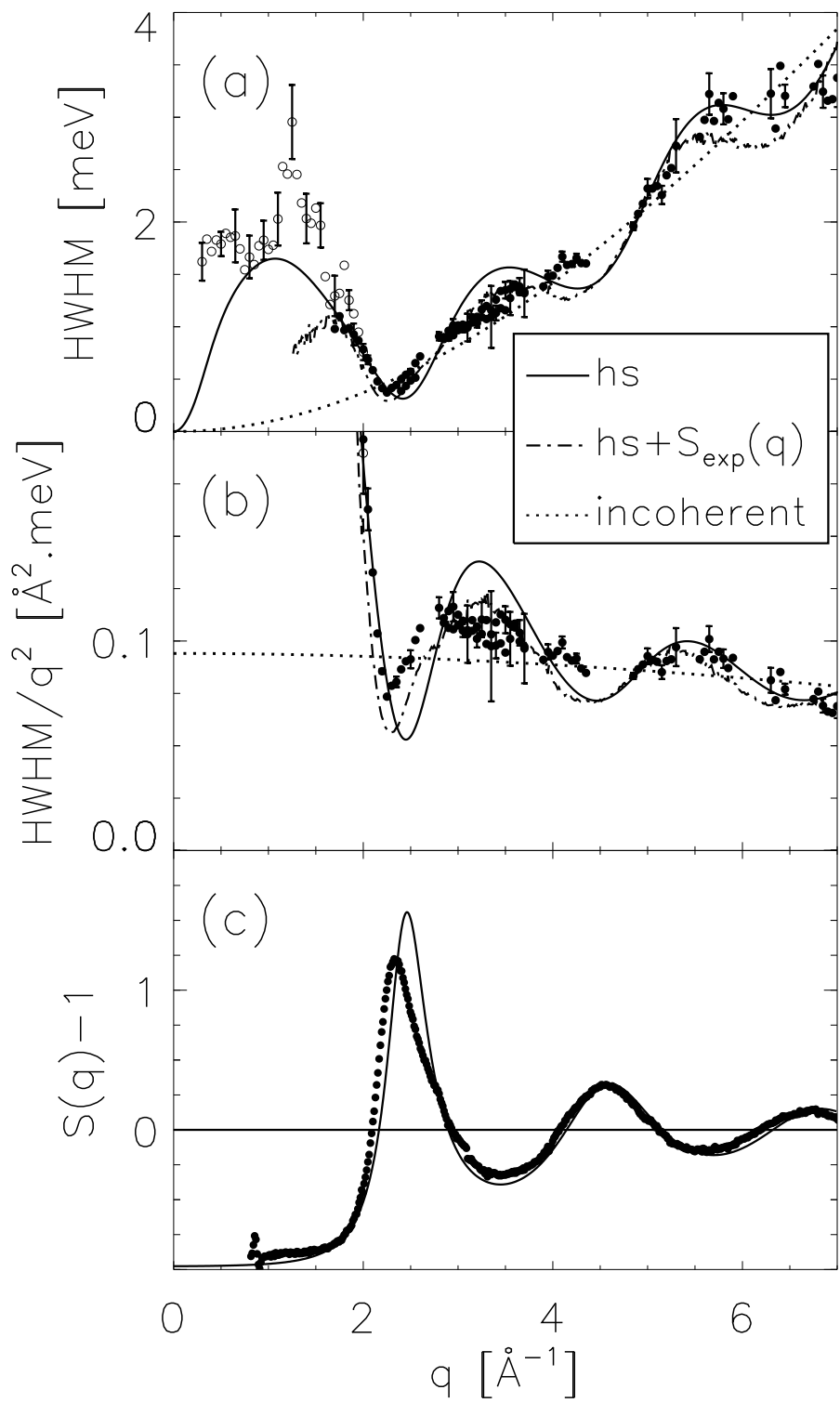
\end{document}